\documentclass[aps,pre,twocolumn,superscriptaddress]{revtex4}
\usepackage{amssymb}
\usepackage{bbm}
\usepackage{indentfirst}
\usepackage{graphicx}
\usepackage{amsmath,amssymb}
\usepackage{bpchem}

\usepackage{color}
\usepackage{tabularx}
\usepackage{multirow}
\usepackage{makecell}
\usepackage{array}
\usepackage{appendix}
\newcommand{\PreserveBackslash}[1]{\let\temp=\\#1\let\\=\temp}
\newcolumntype{C}[1]{>{\PreserveBackslash\centering}p{#1}}
\newcolumntype{R}[1]{>{\PreserveBackslash\raggedleft}p{#1}}
\newcolumntype{L}[1]{>{\PreserveBackslash\raggedright}p{#1}}
\begin{document}

\title{Bifurcation analysis and structural stability of simplicial oscillator populations}

\author{Can Xu}
\email[]{xucan@hqu.edu.cn}
\affiliation{Institute of Systems Science and College of Information Science and Engineering, Huaqiao University, Xiamen 361021, China}

\author{Xuebin Wang}
\affiliation{Institute of Systems Science and College of Information Science and Engineering, Huaqiao University, Xiamen 361021, China}

\author{Per Sebastian Skardal}
\email[]{persebastian.skardal@trincoll.edu}
\affiliation{Department of Mathematics, Trinity College, Hartford, Connecticut 06106, USA }
%

%


\newcommand{\WARN}[1]{\textcolor{green}{#1}}
\newcommand{\NOTES}[1]{\textcolor{red}{#1}}
\begin{abstract}
We present an analytical description for the collective dynamics of oscillator ensembles with higher-order coupling encoded by simplicial structure, which serves as an illustrative and insightful paradigm for brain function and information storage. The novel dynamics of the system, including abrupt desynchronization and multistability, are rigorously characterized and the critical points that correspond to a continuum of first-order phase transitions are found to satisfy universal scaling properties. More importantly, the underlying bifurcation mechanism giving rise to multiple clusters with arbitrary ensemble size is characterized using a rigorous spectral analysis of the stable cluster states. As a consequence of $SO_2$ group symmetry, we show that the continuum of abrupt desynchronization transitions result from the instability of a collective mode under the nontrivial antisymmetric manifold in the high dimensional phase space.
\end{abstract}

\maketitle

\section{Introduction}\label{sec:01}

Spontaneous synchronization in populations of interacting units is a ubiquitous phenomena in complex systems~\cite{pikovsky2001syn,strogatz2003sync}. Describing such collective behaviors in terms of coupled phase oscillators and studying phase transitions between incoherence and synchrony have proven useful in of physics, biology and social systems~\cite{rohden2012self,montbrio2018kuramoto,ottino2018volcano}. Exploring the routes towards synchrony and uncovering the underlying mechanism in various levels have attracted increasing theoretical and experimental interests~\cite{kiss2002emerging,acebron2005the,skardal2015control}.

Most existing literatures focus on pairwise interaction between oscillators that typically contain the first harmonic based on a phase reduction~\cite{kuramoto1975,rodrigues2016the}. However, in many cases, one needs to go beyond such setups allowing for non-pairwise coupling that incorporates high order structures, i.e., simplexes~\cite{salnikov2019simplicial,gong2019low,bibk2016chaos,Ashwin2016PhysD,Leon2019PRE}, which have gained more attention in recent years. Recent advances demonstrate that the simplicial interactions, sometimes called hypernetworks, play essential roles in many systems ranging from signal transmission in neural networks to structural function correlation in brain dynamics~\cite{reimann2017cliques,billings2019simplex}. As shown in~\cite{tanaka2011multi,komarov2015finite,skardal2019abrupt}, notable features of higher-order interactions in coupled oscillator ensembles include the emergence of extensive multistability and a continuum of abrupt desynchronization transitions (ADTs), which lead to potential applications in memory and information storage. Despite these findings, a rigorous analysis of the bifurcations and stability properties that characterize both the macroscopic and microscopic dynamics is lacking, thereby leaving our understanding of these physical phenomena incomplete.

In this paper, we provide an analysis of synchronization transitions induced by simplicial interactions and nonlinear higher order coupling in oscillator ensembles. Using a self-consistent approach we perform a bifurcation analysis that uncovers a general phenomenon that gives rise to ADTs and extensive multistability. Furthermore, we establish scaling relations that describe the critical points for desynchronization that are universal in the sense that they do not depend on the functional form of the natural frequency distribution. Most importantly, we perform a rigorous stability analysis for finitely- and infinitely-many partially synchronized states with arbitrary system size using a spectral analysis of the stable cluster states. We reveal that a continuum of ADTs originate from the instability of a collective mode under the nontrivial antisymmetrical manifold in the high dimensional phase space.

The remainder of this paper is organized as follows. In Sec.~\ref{sec:02} we present a bifurcation analysis for the globally-coupled system. In Sec.~\ref{sec:03} we analyze the stability of cluster states and characterize the properties of the ADTs. In Sec.~\ref{sec:04} we briefly present an additional example of a system with frequency-weighted coupling. Finally, in Sec.~\ref{sec:05} we conclude with a discussion of our results.

\section{Bifurcation Analysis}\label{sec:02}

We consider oscillator ensembles with three-way nonlinear coupling whose evolution is given by
\begin{equation}\label{equ:01}
  \dot{\theta}_i=\omega_i+\frac{1}{N^2}\sum_{m=1}^{N}\sum_{n=1}^{N}K_{imn}\sin(\theta_m+\theta_n-2\theta_i),
\end{equation}
where $\theta_i$ is the phase of oscillator $i$ with $i=1,\ldots,N$, $N$ is the system size, and $\omega_i$ is the natural frequency of oscillator $i$, assumed to be drawn from a distribution $g(\omega)$, which is assumed to be even throughout this paper. Unlike the classical Kuramoto-like models, the coupling term in Eq.~(\ref{equ:01}) is not pairwise, but involves triplets. Finally, $K_{imn}$ is the coupling strength among triplet of oscillators $i$, $m$, and $n$. While we will consider more general cases below, we first restrict our attention to the simplest setup $K_{mni}=K\ge0$ (uniform coupling), in which case Eq.~(\ref{equ:01}) is equivalent to a fully connected hypernetwork topology~\cite{wang2010evolving}. (Later we will consider frequency-weighted coupling.)

Before proceeding with the bifurcation analysis, we introduce two order parameters to characterize the collective behavior of the ensemble, namely, $Z_k=R_k e^{i\Theta_k}=\frac{1}{N} \sum_{m=1}^\infty e^{ik\theta_m}$ for $k=1$ and $2$. In the case where a single synchronized cluster emerges the classical Kuramoto order parameter $Z_1$ is sufficient to describe the macroscopic synchronization dynamics. However, when two clusters emerge the Daido order parameter $Z_2$ measures the degree of (cluster) synchronization~\cite{daido1996multi}, while $Z_1$ measure the degree of asymmetry between the two clusters. $R_{k}$ and $\Theta_{k}$ correspond to amplitudes and arguments, respectively, of the order parameters. By exponentiating the sine term and distributing the summation, these definitions allow us to rewrite Eq.~(\ref{equ:01}) as
\begin{equation}\label{equ:02}
  \dot{\theta}_i=\omega_i+KR_1^2 \sin(2\Theta_1-2\theta_i),
\end{equation}
where $KR_1^2$ acts as an effective force on each oscillator aiming to entrain it via the mean field $\Theta_2$. The presence of a higher order harmonic ($\sim \sin2\theta_i$) and nonlinear coupling ($\sim KR_1^2$), gives rise to a series of nontrivial dynamical features.

Proceeding with our analysis, it is convenient to introduce the parameter $q=KR_1^2$. 
Given the $SO_2$ group symmetry (i.e., rotation and reflection)~\cite{crawford1994amp} of Eq.~(\ref{equ:02}), we enter the appropriate rotating frame, i.e., consider the transformation $\theta\mapsto\theta+\Omega t$ (where $\Omega$ is the mean natural frequency), to obtain a non-rotating solution ($\dot{\Theta}_{k}=0$) and further set $\Theta_{k}=0$ by appropriately shifting initial conditions. In this case, the whole population can be divided into those phase locked oscillators ($|\omega_i|<q$) and that of drifting ones ($|\omega_i|>q$). For the phase locked case, the oscillators are entrained by the mean field, yielding fixed points $\dot{\theta}_i=0$ satisfying
\begin{align}
\sin 2\theta_i=\frac{\omega_i}{q}~~~\text{and}~~~\cos 2\theta_i=\sqrt{1-\frac{\omega_i^2}{q^2}}.\label{equ:02a}
\end{align}
In fact the second-order harmonic in yields two stable fixed points (as well as two unstable fixed points), which manifest in the formation of two clusters with phase difference $\pi$~\cite{dkardal2011cluster,komarov2013multi}. Trigonometric identities allow us to write 
\begin{align}
\sin\theta_i=\frac{\sin 2\theta_i}{2\cos \theta_i}~~~\text{and}~~~\cos\theta_i=\pm\sqrt{\frac{1+\cos 2\theta_i}{2}},\label{equ:02b}
\end{align}
where the $+$ or $-$ in the cosine term determine whether $\theta_i$ falls in the cluster near $\theta=0$ or $\pi$, respectively. Here we replace the $\pm$ in the cosine term with the parameter $p_i$, which takes the value $1$ with probability $\eta(\omega_i)$ and is $-1$ with probability $1-\eta(\omega_i)$. Thus, $\eta(\omega_i)$ is an indicator function satisfying $0\leq\eta(\omega_i)\leq 1$ which represents the fraction of oscillators with natural frequency $\omega_i$ that become entrained to the $\theta=0$ cluster.

Turning briefly to the drifting oscillator population, each oscillator rotates non-uniformly on the unit circle with period $T_i=2\pi/\sqrt{\omega_i^2-q^2}$. Moreover, the contribution of drifiting oscillators to the order parameter $Z_k$ is given by $
\frac{1}{N}\sum_{|\omega_i|>q}\frac{1}{T_i}\int_{0}^{2\pi}\frac{e^{ik\theta_i}}{|\dot{\theta}_i|}d\theta_i$.  In fact, the symmetry of the natural frequency distribution causes this contribution to vanish, implying that drifting oscillators do not contribute to either order parameter $Z_{k}$ whether $N$ is finite or not. The contribution of the locked oscillators to the order parameters can be simplified by first noting that $Z_k=\frac{1}{N} \sum_{m=1}^\infty e^{ik\theta_m}=\frac{1}{N} \sum_{m=1}^\infty \cos(k\theta_m)+i\sin(k\theta_m)$. Similar to the drifting oscillators, the even and odd symmetries of the natural frequency distribution and sine term, respectively, cause the imaginary part to vanish, yielding $R_k=\frac{1}{N} \sum_{m=1}^\infty \cos(k\theta_m)$. Writing these  out explicitly, we have that
\begin{align}
R_1&=\frac{1}{N} \sum_{i=1}^\infty \cos(\theta_i)=\frac{1}{N} \sum_{i=1}^\infty p_i\sqrt{(1+\cos 2\theta_i)/2},\label{equ:03}\\
R_2&=\frac{1}{N} \sum_{i=1}^\infty \cos(2\theta_i)=\frac{1}{N} \sum_{i=1}^\infty \sqrt{1-(\omega_i/q)^2},\label{equ:04}
\end{align}
from which we see that $R_2$ is determined by the parameter $q=KR_1^2$ that controls the range of phase locking, whereas $R_1$ is further restricted by the indicator function $\eta$ that describes the distribution of $p_i$'s, reflecting the asymmetry of phase locked oscillators between two clusters, and therefore is of major importance. Converting sum in Eq.~(\ref{equ:03}) to a sum over frequencies (i.e., treating the population as a distribution) and replacing the cosine term, we find that the parameterized equation for determining $R_1$ is given by
\begin{equation}\label{equ:06}
  \frac{1}{\sqrt{K}}=\frac{1}{N}\sum_{|\omega_i|<q}[2\eta(\omega_i)-1]\sqrt{\frac{1+[1-(\omega_i/q)^2]^{\frac{1}{2}}}{2q}},
\end{equation}
where we denote the right hand side of Eq.~(\ref{equ:06}) as $F(q)$.

\begin{figure*}
  \centering
\footnotesize
  \includegraphics[width=0.55\linewidth]{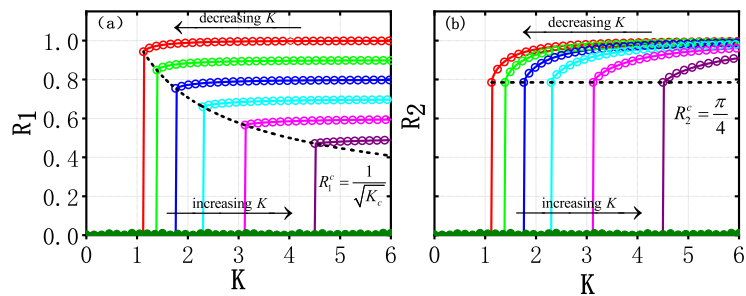}\\
  \caption{Bifurcation diagram of the order parameters $R_1$ (a), $R_2$ (b) with $K$ for the uniform coupling. The circles and solid lines represent the results obtained by numerical simulations and theoretical predictions (stable branches, $q>q_c$), respectively. $\eta=1$ (red), $0.95$ (green), $0.9$ (blue), $0.85$ (cyan), $0.8$ (pink) and $0.75$ (purple). Dashed line denotes the theoretical dependence between $R_1^c$ and $K_c$. For each value of $\eta$, the initial phases are set to $0$ and $\pi$ with probability $\eta$ and $1-\eta$, respectively, then $K$ decreases to $0$ and restores to initial value with $\Delta K=0.01$ adiabatically. In the simulation, $N=10^5$ and $g(\omega)=\frac{1}{2}$, $\omega\in (-1, 1)$.}\label{fig:01}
\end{figure*}

Importantly, Eq.~(\ref{equ:06}) implicitly links $R_{k}$, $K$, and $q$. The underlying bifurcations occur at a critical point $K_c$ given where $\frac{dR_{k}}{dK}|_{K=K_c}=\frac{dR_{k}}{dq}(\frac{dK}{dq})^{-1}|_{q=q_c}$ diverges. Since $F(q)$ is a continuous function we search for either the case $F'(q_c)=0$, which corresponds go a smooth fold bifurcation, or the case where $F'(q_c)$ does not exist, which corresponds to a non-smooth bifurcation. Consequently, the associated critical points are $K_c=F(q_c)^{-2}$ and $R_{k}^c=R_{k}(q_c)$.

To get analytical insights for the critical points, we consider a uniform constant indicator function $\eta(\omega_i)=\eta$. Then $F(q)$ can be simplified to $F(q)=(2\eta-1)f(q)$, where
\begin{align}
f(q)=\frac{1}{N}\sum_{|\omega_i|<q}\sqrt{\frac{1+[1-(\omega_i/q)^2]^{1/2}}{2q}}. \label{equ:06a}
\end{align}
For sufficiently small $K$ no solution exists except for $R_{k}=0$ (which always exists), but when $K$ is increased the bifurcation occurs at the critical value $K_c$ given by
\begin{equation}\label{equ:07}
  K_{c}(\eta)=[(2\eta-1) f(q_c)]^{-2},
\end{equation}
and the corresponding critical order parameters are
\begin{equation}\label{equ:08}
  R_1^c(\eta)=(2\eta-1)\sqrt{q_c}f(q_c),
\end{equation}
\begin{equation}\label{equ:09}
  R_2^c(\eta)=\frac{1}{N}\sum_{|\omega_i|<q_c}\sqrt{1-(\omega_i/q_c)^2}.
\end{equation}

Generically, $q_c$ and $f(q_c)$ are non-zero, indicating a discontinuous jump at $K_c$ between the synchronized (i.e., $R_{k}^c>0$) state and the incoherent (i.e., $R_{k}=0$). Moreover, the incoherent state remains stable for all $K$ in the $N\to\infty$ limit. These two criteria imply that the system undergoes, for a given value of $\eta,$ an ADT characterizing a discontinuous jump from a synchronized state to the incoherent state, but no complementary transition where the incoherent state spontaneously synchronizes. Moreover, a different ADT will occur at each different value of $\eta$, so in fact a continuum of ADTs exist. This phenomenon differs essentially from the traditional first-order phase transition characterized by a finite hysteresis loop~\cite{lee2009large,gomez2011exp} where (i) typically a single desynchronization transition occurs (often via a saddle-node bifurcation) and (ii) a complementary synchronization transition typically follows at a larger value coupling strength.

The self-consistent analysis and scaling formulas presented above capture the macroscopic dynamics exhibited by Eq.~(\ref{equ:01}) with arbitrary $g(\omega)$ and $N$. Note that both $K_c$ and $R_1^c$ exhibit a monotonic dependence on $\eta$ (decreasing and increasing, respectively), whereas $R_2^c$ remarkably remains constant. For instance, for the case of a standard normal frequency distribution, i.e., $g(\omega)=\frac{1}{\sqrt{2\pi}}e^{-\frac{\omega^2}{2}}$~\cite{komarov2015finite}, the ADTs occur at $q_c=1.456$, $f(q_c)=0.679$. For the case of a Lorentzian distribution with unit width, i.e., $g(\omega)=\frac{1}{\pi (\omega^2+1)}$~\cite{skardal2019abrupt}, we have critical points $q_c=1.463$, $f(q_c)=0.491$. Note that for both cases the function $f(q)$ is smooth. However, for the non-smooth and finitely-supported uniform distribution $g(\omega)=\frac{1}{2}$ for $\omega\in (-1,\, 1)$, we have critical points at $q_c=1$ since
\begin{align}
f(q)=\left\{\begin{array}{rl}\frac{2\sqrt{2q}}{3}&\text{if }q<1,\\ \frac{\sqrt{2}\sqrt{q+(q^2-1)^{\frac{1}{2}}}(1+q^2-q\sqrt{q^2-1})}{3q}&\text{if }q\geq1.\end{array}\right.
\end{align}
Moreover, $f'(q_c)$ does not exist since $f'(q_c^+)\neq f'(q_c^-)$, then the corresponding critical points are $K_c=\frac{9}{8(2\eta-1)^2}$ with
\begin{align}
R_1^c=\frac{2\sqrt{2}}{3}(2\eta-1)=\frac{1}{\sqrt{K_c}},~~~\text{and}~~~R_2^c=\frac{\pi}{4}.\label{equ:09a} 
\end{align}
In Fig.~(\ref{fig:01}) we illustrate our results for the uniform distribution, plotting the order parameters $R_1$ and $R_2$ in panels (a) and (b), respectively, using $\eta=1$ (red), $0.95$ (green), $0.9$ (blue), $0.85$ (cyan), $0.8$ (pink), and $0.75$ (purple). Analytical predictions are plotted as solid curves, while results from simulations with $N=10^5$ oscillators are plotted with circles, both as the coupling strength is decreased from a synchronized cluster state and increased from the incoherent state. the analytical prediction matches simulations. Critical values $R_1^c$ and $R_2^c$ are plotted as dotted curves, indicating the continuum of ADTs. Analytical predictions match the simulation results very well.

\section{Stability}\label{sec:03}

To better understand the system dynamics, we consider the structural stability question of what cluster states emerge from an arbitrary configuration. A linear stability analysis demonstrates that the incoherent state can never lose its stability for any $K$. This follows from the fact that the continuous spectrum of the incoherent state, i.e., $R_{k}=0$, is purely imaginary ($i\omega$) and the nonlinear mean field has no contribution to discrete spectrum~\cite{strogatz1991stability}. Thus due to the purely nonlinear interaction term in Eq.~(\ref{equ:02}) it is impossible for a spontaneous phase transition from the incoherent state. We therefore aim to investigate the stability of partially synchronized states, first focusing on finite $N$ case, which then turns out to generalize to the thermodynamic limit $N\to\infty$ directly.

\begin{figure*}
  \centering
  \includegraphics[width=0.55\linewidth]{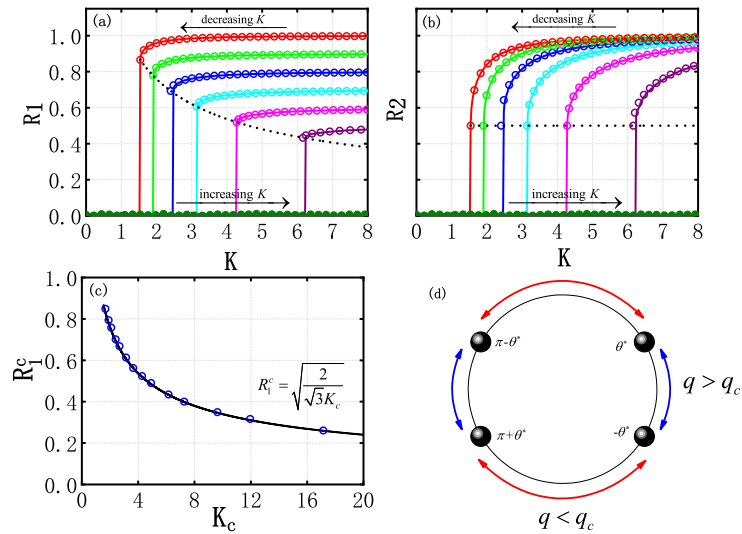}\\
  \caption{Bifurcation diagram of the order parameters $R_1$ (a), $R_2$ (b) with $K$ for the frequency weighted coupling, $g(\omega)=\frac{1}{\pi(\omega^2+1)}$. Circles are numerical simulations and solid lines are theoretical results (stable branches, $q>q_c$). $\eta=1$ (red), $0.95$ (green), $0.9$ (blue), $0.85$ (cyan), $0.8$ (pink) and $0.75$ (purple). (c) Analytical relation between $R_1^c$ and $K_c$ (black solid line) with $g(\omega)=\frac{1}{\sqrt{2\pi}}e^{-\frac{\omega^2}{2}}$, blue circles represent numerical simulations with different $\eta$, which is in agreement with the dashed line (a) representing universal phase transition. (d) Sketch map of the stable configuration, the blue arrows (red) stand for a/an stable (unstable) perturbed direction. Numerical strategy is the same as Fig.~(\ref{fig:01}).}\label{fig:02}
\end{figure*}

The $SO_2$ group symmetry ensures that the drifting oscillators generate a purely imaginary spectrum in the linear stability analysis, thereby having no contributions to the mean field~\cite{omel2012non,omel2013bifurcation}. In other words, they effectively decouple from the locked oscillators in Eq.~(\ref{equ:02}), so we can only consider $N_l$ oscillators locked to the mean field for some $q$ excluding the drifting ones. We note, however, that in some cases, such as uniformly-distributed frequencies, all oscillators become locked in a synchronized state, yielding $N_l=N$~\cite{mirollo2005spe}. To study stability of such partially synchronized states, we consider a small perturbation $x_i$, $|x_i|\ll1$, on each phase locked oscillator $\theta_i$ in Eq.~(\ref{equ:02}) and, after neglecting high order terms, the evolution for perturbation satisfies $\dot{\mathbf{x}}=\mathbf{J}\mathbf{x}$ where $\mathbf{x}=(x_1,\ldots, x_{N_l})$ and $\mathbf{J}$ is the Jacobian matrix with entries $J_{ij}=\frac{\partial \dot{\theta}_i}{\partial \theta_j}$. In fact, $\mathbf{J}$ is of the form
\begin{equation}\label{equ:13}
  \mathbf{J}=\frac{2KR_1}{N}\mathbf{M}-2q\mathbf{D},
\end{equation}
where the matrix $\mathbf{M}$ is given by $M_{ij}=\cos(\theta_j-2\theta_i)$ and a diagonal matrix $\mathbf{D}$ is given by $D_{ij}=\cos(2\theta_i)\delta_{ij}$. The stability properties for a given cluster state is then determined by the eigenvalues of $\mathbf{J}$. First, note that $\sum_{j=1}^{N_l} J_{ij}=0$, which gives a trivial eigenvalue $\lambda=0$ (with constant eigenvector $\mathbf{v}\propto\mathbf{1}=(1,\ldots,1)$) that corresponds to the rotation invariance of Eqs.~(\ref{equ:01}) and (\ref{equ:02}). The remaining $N_l-1$ eigenvalues can be calculated via the characteristic equation of $\mathbf{J}$, which can be expressed as $B(\lambda)=\text{det}(\lambda\mathbf{I}-\mathbf{J})$. Next, by defining
\begin{align}
\mathbf{E}=\frac{\lambda \mathbf{I}}{2K R_1}+R_1\mathbf{D},
\end{align}
one may write
\begin{align}
2KR_1\mathbf{E}(\mathbf{I}-\mathbf{E}^{-1}\mathbf{M}/N).
\end{align}
Thus, the characteristic polynomial takes the form
\begin{align}
B(\lambda)=2KR_1 \text{det}(\mathbf{E})\text{det}(\mathbf{I}-\frac{1}{N}\mathbf{E}^{-1}\mathbf{M}).
\end{align}
Finally, assuming that $\mathbf{E}$ must is in fact invertible, we have that $det(\mathbf{E})\neq 0$, and the key task for calculating the eigenvalue spectrum depends on finding the roots of the last term alone.

Next we ease notation by defining four vectors $\mathbf{c}^{(k)}$ with $c_i^{(k)}=\cos k\theta_i$ and $\mathbf{s}^{(k)}$ with $s_{i}^{(k)}=\sin k\theta_i$ that satisfy the orthogonality property $\mathbf{c}^{(k)}\cdot \mathbf{s}^{(k')}=0$ ($k'=1,2$). We note here that $k=2$ corresponds to purely deterministic vectors, while $k=1$ is random vectors due to clustering and multi-branches. That is, the entries of $\mathbf{c}^{(2)}$ and $\mathbf{s}^{(2)}$ do not depend on whether a given oscillator falls in the $0$ or $\pi$ cluster, but the entries of $\mathbf{c}^{(1)}$ and $\mathbf{s}^{(1)}$ do. Moreover, we note that the rank of $\mathbf{M}$ is only $2$ since, for all $\mathbf{x}\in \mathbb{R}^{N_l}$, we have $\mathbf{M}\mathbf{x}=(\mathbf{c}^{(1)}\cdot\mathbf{x})\mathbf{c}^{(2)}+(\mathbf{s}^{(1)}\cdot\mathbf{x})\mathbf{s}^{(2)}$. Introducing the orthonormal basis $\mathbf{a}_1=\frac{\mathbf{c}^{(1)}}{\|\mathbf{c}^{(1)}\|}$ and $\mathbf{a}_2=\frac{\mathbf{s}^{(1)}}{\|\mathbf{s}^{(1)}\|}$, the matrix $\mathbf{E}^{-1}\mathbf{M}/N$ restricted to the subspace that is spanned by $\{\mathbf{a}_{i}\}$ turns out to be a $2\times 2$ matrix $\mathbf{Q}$, i.e., $(\mathbf{E}^{-1}\mathbf{M}/N)_{\mathbf{a}_1,\mathbf{a}_2}=\mathbf{Q}(\lambda)$.
The entries of this matrix are defined as $Q_{ij}=\mathbf{a}_i\cdot\frac{1}{N}\mathbf{E}^{-1}\mathbf{M}\mathbf{a}_j$, whose diagonal entries are
\begin{equation}\label{equ:17}
  Q_{11}=\frac{1}{N}\sum_{i=1}^{N_l} [2\eta(\omega_i)-1]\frac{2KR_1 |c_i^{(1)}| c_i^{(2)}}{\lambda+2qc_i^{(2)}},
\end{equation}
\begin{equation}\label{equ:20}
  Q_{22}=\frac{1}{N}\sum_{i=1}^{N_l} [2\eta(\omega_i)-1]\frac{KR_1 (s_i^{(2)})^2/ |c_i^{(1)}|}{\lambda+2qc_i^{(2)}},
\end{equation}
and the off-diagonal $Q_{ij}\propto \frac{1}{N}\sum_{m=1}^{N_l} \frac{c_m^{(i)}s_m^{(j)}}{E_{mm}}$ with $i\neq j$, while the other $N_l-2$ basis vectors span the kernel of $\mathbf{E}^{-1}\mathbf{M}/N$. Orthogonality implies that $Q_{12}$ and $Q_{21}$ are zero, so the characteristic polynomial simplifies to
\begin{equation}\label{equ:21}
  B(\lambda)=2KR_1 \prod_{i=1}^{N_l} E_{ii}[1-Q_{11}(\lambda)][1-Q_{22}(\lambda)].
\end{equation}

In fact, the rational functions $Q_{jj}(\lambda)$ ($j=1, 2$) are strictly decreasing away from their $\frac{N_l}{2}$ poles $\lambda_i=-2qc_i^{(2)}$ ($\lim_{\lambda\to \lambda_i^\pm}Q_{jj}(\lambda)=\pm\infty$), so $Q_{jj}(\lambda)=1$ must have one route between two consecutive poles. The necessary condition for stable locked state requires all $c_i^{(2)}>0$. In addition, we find that $Q_{11}(0)=1$ corresponds to a trivial eigenvalue $\lambda=0$. The remaining one eigenvalue is uniquely determined by $Q_{22}(\lambda)=1$ for $\lambda>\lambda_i$. Since $Q_{22}(\lambda)$ is decreasing, the root is negative if and only if $Q_{22}(0)<1$, which leads to the structural stability condition for the configuration of population
\begin{equation}\label{equ:22}
  \frac{1}{N}\sum_{i=1}^{N_l} [2\eta(\omega_i)-1]\frac{KR_1 (s_i^{(2)})^2}{2q|c_i^{(1)}|c_i^{(2)}}<1,
\end{equation}
or equivalently, $F'(q)<0$. Therefore, we conclude that the eigen-spectrum of $\mathbf{J}$ is made up of three parts, a heavily populated part consisting of $N_l-2$ negative eigenvalues within the poles, a trivial part $\lambda=0$ corresponding to rotation invariance, and a lone eigenvalue outside the poles. In the thermodynamic limit $N\to\infty$, the first part merges into continuous spectrum with $\lambda(\omega)=-2\sqrt{q^2-\omega^2}$, while the last part turns out to be a nontrivial discrete spectrum determined by the continuum limit $Q_{22}(\lambda)=1$ ($\lambda \neq \lambda(\omega))$.

It is also instructive to characterize the eigenvectors of $\mathbf{J}$. Considering a frequency-dependent vector $\mathbf{x}\in\mathbb{R}^{N_l}$ with $x_i=\chi(\omega_i)$, the space $\mathbb{R}^{N_l}$ can be split into two subspaces $V_{even}$ and $V_{odd}$. If $\chi$ is an even (or odd) function, then $\mathbf{x}\in V_{even}$ (or $V_{odd}$). This definition allows $\mathbf{x}$ to be random vector, such as $\mathbf{c}^{(k)}\in V_{even}$, $\mathbf{s}^{(k)}\in V_{odd}$, so $V_{even}\perp V_{odd}$ and $V=V_{even}\oplus V_{odd}$. The eigenspace of $\mathbf{J}$ can be described via the basis of $V_{even}$ and $V_{odd}$. For $\mathbf{x}^e\in V_{even}$, we have $\mathbf{J}\mathbf{x}^e=\frac{2KR_1}{N}\mathbf{c}^{(2)}(\mathbf{c}^{(1)}\cdot\mathbf{x}^e)-2q\mathbf{D}\mathbf{x}^e$. Setting $\mathbf{c}^{(1)}\cdot \mathbf{x}^e=1$, the eigenvector equation $\mathbf{J}\mathbf{x}^e=\lambda\mathbf{x}^e$ implies that $x_i^e=c_i^{(2)}/[NE_{ii}(\lambda)]$ in which $\lambda$ is a root of $Q_{11}(\lambda)=1$. Similarly, if $\mathbf{x}^o\in V_{odd}$, imposing $\mathbf{s}^{(1)}\cdot\mathbf{x}^o=1$, we obtain $x_i^o=s_i^{(2)}/[NE_{ii}(\lambda)]$ corresponding to $Q_{22}(\lambda)=1$. It is worth mentioning that $\mathrm{Re}[\nabla (Z_k)\cdot \mathbf{x}^{e}]=0$ and $\mathrm{Im}[\nabla(Z_k)\cdot\mathbf{x}^o]= 0$. 

The structural stability of the clusters can then be interpreted as follows. Recall that, after entering a rotating frame and shifting initial conditions we have the mean phases $\Theta_{k}=0$. This corresponds to two clusters: one centered at $\theta=0$ and the other at $\theta=\pi$. Preserving this particular symmetry, perturbation may then contract or spread oscillators in each cluster, effectively pulling oscillators along the unit circle in the real or imaginary directions, respectively. Our analysis above reveals that the eigenvalues determined by $Q_{11}(\lambda)=1$ correspond to even eigenvectors, i.e., symmetric perturbation, that induce purely imaginary displacement of the centroid of configuration in the linear approximation, thereby inducing a small spread in each cluster. On the other hand, the eigenvalues for $Q_{22}(\lambda)=1$ correspond to odd eigenvectors, i.e., antisymmetric perturbation, that leads to purely real displacement of the centroid, thereby contracting each cluster. As the structure parameter $q$ is decreasing, the nontrivial solitary eigenvalue is shifted towards positive real axis and collides with the origin at $q=q_c$. Thus, the continuum of ADTs originate from the instability of a collective mode under the nontrivial antisymmetric perturbation in the high dimensional phase space associated with this solitary eigenvalue.

\section{Another Example: Frequency-Weighted Coupling}\label{sec:04}

Lastly, we show that our theory generalizes by considering the coupling heterogeneity of the form $K_{mni}=K|\omega_i|$, i.e., establishing the correlation between oscillators frequency and coupling strength as in Refs.~\cite{zhang2013exp,bi2016coe} that generates explosive synchronization and quantized time-dependent clustering. We show that such a frequency-weighted coupling scheme captures the essential properties of simplicial dynamics and can achieve certain cluster arrangement. In this case, the $SO_2$ group symmetry still holds, and the mean-field equation can be written as
\begin{align}
\dot{\theta}_i=\omega_i-q|\omega_i|\sin(2\theta_i).
\end{align}
If $q<1$, no oscillators are entrained by the mean-field indicating the asynchronous state. Interestingly, once $q\geq 1$, all oscillators become phase-locked ($N_l=N$) with $c_i^{(2)f}=\sqrt{1-q^{-2}}$ and $s_i^{(2)f}=\omega_i/(q|\omega_i|)$. The parametric function $F(q)$ is a smooth function for a constant $\eta$, i.e.,
\begin{align}
F(q)=(2\eta-1)\sqrt{[1+(1-q^{-2})^{\frac{1}{2}}]/(2q)},
\end{align}
and the fold bifurcation characterizing a continuum of universal ADTs occurs at $q_c=2/\sqrt{3}$ with
\begin{align}
K_c=\frac{8}{3\sqrt{3}(2\eta-1)^2},~~R_1^c=\frac{\sqrt{3}(2\eta-1)}{2},~~\text{and}~~R_2^c=\frac{1}{2}. 
\end{align}
The results are presented in Fig.~\ref{fig:02} (a)--(c) (details similar to those for Fig.~\ref{fig:01}), displaying remarkable agreement between theory and simulation.

The stability analysis for the multi-clusters follows the similar way, where the Jacobian matrix is
\begin{align}
\mathbf{J}^f=\mathbf{W}(\frac{2KR_1}{N}\mathbf{M}-2q\mathbf{D})
\end{align}
with the frequency matrix $\mathbf{W}$ being $W_{ij}=|\omega_i|\delta_{ij}$. The resulting characteristic polynomial is
\begin{align}
B^f(\lambda)=2KR_1 \text{det}(\mathbf{E}^f)[1-Q_{11}^f(\lambda)][1-Q_{22}^f(\lambda)],
\end{align}
where
\begin{align}
\mathbf{E}^f&=\frac{\mathbf{W}^{-1}\lambda\mathbf{I}}{2KR_1}+R_1\mathbf{D},\\
Q_{11}^f&=\frac{1}{N}\sum_{i=1}^{N}\frac{c_i^{(1)f} c_i^{(2)f}}{E_{ii}^f(\lambda)},~~\text{and}\\
Q_{22}^f&=\frac{1}{N}\sum_{i=1}^{N}\frac{s_i^{(1)f} s_i^{(2)f}}{E_{ii}^f(\lambda)}.
\end{align}
The eigenvalue spectrum together with the associated eigenvectors has the same appearance as uniform coupling. In particular, the phase locked states exhibit fixed cluster arrangement regardless of $g(\omega)$, namely, the oscillators are recruited into two symmetric groups $\pm\theta^*$ ($\eta=1$) or four groups $\pm\theta^*$, $\pi\pm\theta^*$ ($\eta<1$). These different clusters can be understood as distinct picks in the information storage, while the stable configuration for the multi-clusters $Q_{22}^f(0)<1$ is equivalent to $\tan\theta^*\cdot \tan 2\theta^*<1$ impling $\theta^*<\frac{\pi}{6}$. Decreasing $q$ makes the symmetric clusters move away from each other (Fig.~(\ref{fig:02})d) and the configuration becomes unstable once $q<q_c$. The ADTs occur owing to a saddle node bifurcation where all oscillators become unlocked corresponding to the resting state.

\section{Discussion}\label{sec:05}

Summarizing, we have presented rigorous analytical descriptions for the collective dynamics in a population of globally coupled phase oscillators with higher-order interactions via simplicial structure. Extensive multistability of clusters and ADTs emerge directly from these higher order interactions and the nonlinear mean field. These results were obtained using a self-consistency analysis and rigorous bifurcation theory and reveal scaling properties of the transitions in the form of dependence of the critical points on the indicator and structural constants. The rigorous characterization of the eigenvalue spectrum presented above demonstrates that the stability properties of infinite many partially synchronized states are controlled by a nontrivial solitary eigenvalue, while the occurrences of ADTs correspond to the instability of a collective mode.

\section*{ACKNOWLEDGEMENTS}

C. X. and X. W. acknowledge support from the National Natural Science Foundation of China (Grants Nos. 11905068 and 11847013).

\end{document}